\newcount\nummer \nummer=0
 \def\f#1{\global\advance\nummer by 1 \eqno{(\number\nummer)}
       \global\edef#1{(\number\nummer)}}
 
 \def\be{\begin{equation}}
 \def\ee{\end{equation}}
 \def\eel#1 {\label{#1}\end{equation}}
 \def\ba{\begin{eqnarray}}
 \def\ea{\end{eqnarray}}
 
 \def\re{(\ref}
 \def\rz#1 {(\ref{#1}) } \def\ry#1 {(\ref{#1})}
 
 \let\a=\alpha \let\b=\beta \let\g=\gamma \let\d=\delta

   \let\l=\lambda \let\m=\mu
 \let\n=\nu    \let\s=\sigma
  \let\o=\omega  
    
 \let\O=\Omega \let\S=\Sigma

 \def\0{\over } \def\1{\vec } \def\2{{1\over2}} \def\4{{1\over4}}
 \def\5{\bar } \def\6{\partial }
 \def\7#1{{#1}\llap{/}}
 \def\8#1{{\textstyle{#1}}} \def\9#1{{\bf {#1}}}

 \def\({\left(} \def\){\right)} \def\<{\langle } \def\>{\rangle }
 \def\[{\left[} \def\]{\right]}  
 \let\lra=\leftrightarrow 
  \let\ra=\rightarrow

         \let\ti=\tilde

 \newcommand{\dR}{\mbox{{\sl I \hspace{-0.8em} R}}}
 \def\gi{g^{-1}} \def\gd{g_\downarrow} \def\gu{g_\uparrow}
 \def\gdi{g_\downarrow^{-1}} \def\gui{g_\uparrow^{-1}}

 \def\ua#1.{^{#1}_\uparrow} \def\da#1.{^{#1}_\downarrow}
\def\uA{_\uparrow} \def\dA{_\downarrow}

\documentstyle[11pt]{article}

\begin{document}

\begin{titlepage}
\vspace{-2cm}
\title{{\bf The Topological  G/G WZW Model
in the Generalized Momentum Representation}}
\author{\\ \\
{\bf ANTON YU. ALEKSEEV}\thanks{On leave of
absence from Steklov Mathematical Institute,
Fontanka 27, St.Petersburg, Russia; \newline  e-mail:
alekseev@itp.phys.ethz.ch \newline
\hspace*{5 pt}
$^\heartsuit$e-mail:schaller@tph.tuwien.ac.at \newline
\hspace*{2pt}
$^\diamondsuit$e-mail: tstrobl@pluto.physik.rwth-aachen.de}
\\
Institut f\"{u}r Theoretische Physik, ETH-H\"{o}nggerberg
\\
CH-8093 Z\"{u}rich, Switzerland
\\[4mm]
{\bf PETER SCHALLER}$^\heartsuit$
\\
Institut f\"{u}r Theoretische Physik, TU Wien
\\
Wiedner Hauptstr. 8-10, A-1040 Vienna, Austria
\\[4mm]
{\bf THOMAS STROBL}$^\diamondsuit$
\\
Institut f\"{u}r Theoretische Physik, RWTH Aachen
\\
Sommerfeldstr. 14, D-52056 Aachen, Germany
\\ \\ }

\vspace{0.6cm}
\date{May  1995}
\vfill
\maketitle \thispagestyle{empty}

\vspace{0.5cm}
\begin{abstract}
We consider the topological gauged WZW model in
the generalized momentum representation. The chiral field $g$
is interpreted as a counterpart of the electric
field $E$ of conventional gauge theories.
The gauge dependence of wave
functionals $\Psi(g)$ is governed by a new gauge
cocycle $\phi_{GWZW}$. We evaluate this cocycle
explicitly using the machinery of Poisson $\sigma$-models.
In this approach the GWZW model is reformulated as a
Schwarz type topological theory so that the action
does not depend on the world-sheet metric.
The equivalence of this new formulation to the original
one is proved for genus one and conjectured for
an arbitrary genus Riemann surface. As a by-product
we discover a new way to explain the appearance of
Quantum Groups in the WZW model.
\end{abstract}
\vspace*{-19.5cm}
\hspace*{9cm}
{\large \tt  ETH-TH/95-14 \ \ TUW-95-07} \\
\hspace*{9cm}
{\large \tt PITHA - 95/9 \  ESI$\,$225$\,$(1995)}         \\
\hspace*{9cm}
{\large \tt hep-th/9505012}
\end{titlepage}

\section*{Introduction}
In this paper we investigate the quantization of the gauged $(G/G)$
WZW model in the generalized momentum representation.
The consideration is
inspired by the  study of (two-dimensional) Yang-Mills and
BF-theories in the  momentum representation \cite{AER}.

The problem of quantization of gauge theories in the momentum
representation has been attracting attention for a long time
\cite{JG}.\footnote{The authors
are grateful to Prof. R.Jackiw for drawing their attention to this
paper and for making them
know about  the scientific content of his letter to Prof. D. Amati.}
This question arises naturally in the Hamiltonian version
of the functional intergral formalism \cite{FRS}.
While in the connection representation the idea of gauge
invariance may be implemented in a simple way
\be
\Psi(A^g)=\Psi(A) \ ,
\eel PsiA
we get a nontrivial
behavior of the quantum wave functions
under gauge transformations in the momentum representation.
Indeed,
one can apply the following simple argument.
The wave functional in the momentum representation may be thought of as a
functional Fourier transformation of  the wave functional in
the connection representation (\ref{PsiA}):
\be
\Psi(E)=\int {\cal D}A \,  \exp \left(i \int tr E_i A_i \right) \,  \Psi(A) \ .
\eel A-E
Taking into account the behavior of  $A$ and $E$
under gauge transformations,
\begin{eqnarray}  \label{trans}
A_i^g&=&g^{-1}A_i g+g^{-1}\partial_i g \, ,  \nonumber   \\
E_{i}^g&=&g^{-1}E_i g \,  ,
\end{eqnarray}
we derive
\be
\Psi(E^g)= e^{i\int tr E_i \partial_igg^{-1}} \Psi(E) \, .
\eel PsiE
We conclude that the wave functional in the momentum representation is
not invariant with respect to gauge transformations. Instead, it gains
a simple phase factor $\phi(E, g)$, which is of the form
\be
\phi(E, g)=\int tr E_i \partial_igg^{-1} .
\eel phi
The infinitesimal version of the same phase factor,
\be
\phi(E, \epsilon)=\int tr E_i \partial_i \epsilon \, ,
\eel ephi
corresponds to the action of the gauge algebra.

It is easy to verify that $\phi$ satisfies the following equation
\be
\phi(E, gh) =  \phi(E, g) + \phi(E^g, h) \, .
\eel cocycle
This property assures that the composition of two gauge
transformations \rz PsiE with gauge parameters $g$ and $h$ is the
same as a gauge transformation with a parameter $gh$. Eq.\ \rz
cocycle is usually referred to as a cocycle condition. It
establishes the fact that $\phi$ is a one-cocycle of the (infinite
dimensional) gauge group.
A one-cocycle is said to be trivial, if there exists  some
$\tilde\phi$ such that:
\be
 \phi(E, g) =\tilde\phi (E^g)-\tilde\phi(E)  \, .
\eel cobcon
 In this case the gauge invariance of the wave
function may be restored by the redefinition
\be
\tilde{\Psi}(E)=e^{-i\tilde{\phi}(E)}\Psi(E).
\eel tilPsi

An infinitesimal cocycle $\phi(E, \epsilon)$ is trivial,
if it can be represented as a function of the commutator $[\epsilon, E]$:
\be
\phi(E, \epsilon)=\tilde\phi ([\epsilon, E]).
\eel ecob
It is easy to see that the cocycle \rz ephi is nontrivial.
Indeed, let us choose both $E$ and $\epsilon$ having only one
nonzero component $E^a$ and $\epsilon^a$ (in the Lie algebra).
Then the commutator
in \rz ecob is always equal to zero, whereas the expression
\rz ephi is still nontrivial. As a consequence, also the gauge group cocycle
\rz phi is nontrivial.

On the other hand, on some {\em restricted} space of values for
the field $E$ the cocycle may become trivial
(generically if we admit nonlocal expressions for $\tilde{\phi}$).
This is important to
mention as one may rewrite the (integrated) Gauss law  (\ref{PsiE}) as
a triviality condition on the cocycle:  Let us parameterize  $\Psi$ as
$\Psi(E) :=\exp i\tilde\phi(E)$, which is  possible whenever $\Psi \neq 0$,
and insert this expression into (\ref{PsiE}). The result is
precisely (\ref{cobcon})
with $\phi \equiv \int tr E_i \partial_igg^{-1}$.\footnote{More accurately,
one obtains (\ref{cobcon}) only  mod $2\pi$. But anyway this
modification of (\ref{cobcon}) is quite
natural in view of the origin of the cocycle within
\ry PsiE . Alternatively one might
regard also a multiplicative cocycle $\Phi=\exp (i\phi)$
right from the outset,
cf.\ Appendix A.} In fact, e.g.\ in
two dimensions the wave functions of the momentum representation
are supported on  some conjugacy classes $E(x)=g(x)\,E_0\,g^{-1}(x)$
with specific
values of $E_0$. But away from these specific conjugacy classes, and
in particular in the original, unrestricted space of values for
$E$, the general argument of the cocycle \rz cobcon being nontrivial applies.
More details on this issue may be found in Appendix A.

It is worth mentioning that in the Chern-Simons theory
a cocycle  appears in the
connection representation as well:
\be
\Psi(A^g)=e^{i\phi(A, g)}\Psi(A).
\eel anomaly
The cocycle $\phi(A, g)$ is usually called Wess-Zumino action
\cite{Wess}. It is intimately related to the theory of
anomalies \cite{FSh}.

Recently, a cocycle of type \rz phi has been observed in
two-dimensional BF- and YM-theories.
In this paper we
consider the somewhat more complicated example of
the gauged WZW (GWZW) model for a  semi-simple  Lie group.
Like the BF theory, it is a two-dimensional
topological field theory \cite{Sp} (for a detailed account
see \cite{Blau}).
It has a connection one-form (gauge field)
as one of its dynamical variables and possesses the usual
gauge symmetry. However, there is a complication which
makes the analysis different from the  pattern \re{PsiE}).
In the GWZW model the variable  which is conjugate to the gauge
field, and which shall be denoted by $g$ in the
following,  takes values in a Lie group $G$ instead of a linear space.  So, we
get a sort of curved momentum space. We calculate
the cocycle $\phi_{GWZW}$ which governs the gauge dependence
of wave functions in a $g$-representation and find that it differs
from the standard expression \re{phi}). We argue that while the cocycle \rz phi
corresponds to a Lie group $G$, our cocycle is related to its quantum
deformation $G_q$.
In the course of the analysis we find that the  GWZW
model belongs to the class of Poisson $\sigma$-models recently
discovered in \cite{Th}. This theory provides a technical tool for
the evaluation of the cocycle $\phi_{GWZW}$.

Let us briefly characterize the content of each section. In
Section 1 we develop the Hamiltonian formulation of the GWZW model, find
canonically conjugate variables, and  write down the gauge invariance
equation for the wave functional in the $g$-representation.

Section 2 is devoted to the description of Poisson $\sigma$-models.
A two-dimensional topological $\sigma$-model of this class is
defined by fixing a Poisson bracket on the target space.
Using the Hamiltonian formulation (the topology of the space-time
being a torus or cylinder), we prove that the GWZW model is equivalent
to a certain Poisson $\sigma$-model
coupled to a `topological' term $S_\d$ that has support of measure
zero on the target space of the field theory.
The target space of the (coupled)
Poisson $\sigma$-model is the Lie group $G$.
We start from the GWZW action, evaluate
the Poisson structure on $G$ and discover its relation to
the theory of Quantum Groups. The origin of the term $S_\d$ is
considered in details in Appendix B.

In Section 3 we solve the gauge
invariance equation and find the gauge dependence of the GWZW wave
functional in the $g$-representation. This provides
a new cocycle $\phi_{GWZW}$.
Calculations are performed for the Poisson $\sigma$-model
without the singular term.
In Appendix C we reconsider the problem in the presence
of the topological term. It is shown that in the case
of $G=SU(2)$ at most one quantum state is affected.
We compare the results with other approaches
\cite{ADW}, \cite{Blau}.

In some final remarks we conjecture that the Poisson
$\sigma$-model coupled to $S_\d$
gives an alternative formulation of the GWZW model
valid for a Riemann surface of arbitrary genus. We comment
on the new relation between the WZW model and Quantum
Groups which emerges as a by-product of our consideration.

 \section{Hamiltonian Formulation of the GWZW Model}
 The WZW theory is defined by the action
 \be
   WZW(g)={ k\over 8\pi} \int  tr \, \6_\mu gg^{-1} \6^\mu gg^{-1}\, d^2x +
          {k \0 12\pi} \int\, tr \, d^{-1}(dgg^{-1})^3 \, ,
 \eel acwzw
where  the fields $g$ take values in some semi-simple Lie group $G$ and
indices $\mu$ are raised with the standard Minkowski  metric.
The case of a Euclidean metric may be treated in the same fashion.
Some remarks concerning the second term in \rz acwzw may be found in
Appendix B.

The simplest way to gauge
 the global symmetry transformations $g\to lgl^{-1}$ is to
introduce a gauge field $h$ taking its values in the gauge group; the
action
\be GWZW(h,g)=WZW(hgh^{-1}) \ee
is then  invariant under the local
 transformations $g\to lgl^{-1}, h \to hl^{-1}$. With the celebrated
 Polyakov-Wigmann formula and  $a_\pm :=h^{-1}
\6_\pm h$, where $\6_\pm =\6_o \pm\6_1$,
GWZW can be brought into the  standard form
 \be
  \begin{array}{r@{}l}
    GWZW&(g,a_+,a_-)= WZW(g) \,  + \\[2pt]
         &+ \, {k\over 4\pi} \int tr [a_+\6_-g\gi-a_-\gi \6_+g -a_+ga_-\gi
                     +a_+a_- ] \, d^2x \, .
  \end{array}
 \eel agwzw
In the course of our  construction of GWZW $\;$
$a \equiv a_+ dx^+ +  a_- dx^-$ has been
subject to the zero curvature condition $da + a^2  \equiv 0$.
This condition results also from the   equations of motion arising from
(\ref{agwzw}). So, further on we treat $a_\pm$  as unconstrained fields
(taking their values in the Lie  algebra of the chosen gauge group).

In order to find a Hamiltonian formulation of the GWZW model, we first
 bring \rz agwzw into   first order form. To this end we
introduce an auxiliary field $p(x)$ into the action by the replacement
($\dot g \equiv \6_0 g$)
\be {k \0 8\pi} (\dot g g^{-1} + a_+ - g a_- \gi )^2
 \to \, p (\dot g g^{-1} + a_+ - g a_- \gi ) -  {2\pi \0 k} p^2 \, .
\eel curre
As  $p$ enters  the action quadratically, it may be
 eliminated always  by means of its equations of motion so as to
reproduce the original action (\ref{agwzw}).
 In the functional integral
approach this corresponds to performing the Gaussian integral over $p$.

Now the action (\ref{agwzw})
may be seen to take the form (with $\partial g \equiv \partial_1 g$)
\begin{eqnarray} \label{fofac}
&GWZW\!\!\!\!&
 (g, p, a_\pm)= {k \0 12\pi} \int\, tr d^{-1}(dgg^{-1})^3 \, +
\int d^2x\, tr \Biggl\{ p \dot g\gi - \nonumber  \\
&&- a_-\left[\gi pg-p+  {k\over 4\pi}(\gi
\6 g+\6 g\gi)\right]-\\&&
 - p \6g\gi -
 {k\over 8\pi} (a_+ - a_- - {4\pi \0 k} p +
\6 g \gi )^2 \Biggr\}  \, .
\nonumber
\end{eqnarray}
This is already linear in time derivatives.
After the simple shift of variables
\be a_+ \to \ti a_+ \equiv a_+ - a_- - {4\pi \0 k} p +
\6 g \gi  \ee
the last term is seen to completely decouple from the rest of the
action. Therefore one can  exclude it from the action without loss of
information.  We can again employ  the argument about integration
over $\ti a_+$ (or also $a_+$ in \re{fofac})). So we have introduced
one extra variable $p$ and now  one variable is found to
drop out from the formalism.

After $\ti a_+$ is excluded, the rest of  formula (\ref{fofac}) provides
the Hamiltonian formulation of the model.
The first two terms play the role of a symplectic potential,
giving rise to the symplectic form\footnote{Cf.\ also Appendix C.}
 \be
 \O^{field}=tr \oint \left[ dpd g \gi +
                \left(p  + {k\over 4\pi}\6 g\gi\right)
 \left(d g \gi\right)^2 \right]dx^1     \; .
  \eel psss
Here $d$ is interpreted as an exterior derivative on the phase space.
It is interesting to note that the nonlocal  term in (\ref{fofac})
gives a local contribution to the symplectic form on the phase space.
The third term, which includes
$a_-$, represents a constraint:
 \be
   \gi  p  g- p+{k\over 4\pi} \left(\gi \6 g+\6 g\gi\right) \approx 0.
 \eel const
The variable $a_-$ is a Lagrange multiplier and the constraint is
nothing but the Gauss law of the GWZW model. It is a nice exercise to
check with the help of
(\ref{psss})
that the constraints \rz const are first class and that they
generate the gauge transformations.
Equation \rz const implies
$tr \left(\gi p g + {k\over 4\pi}\gi \6 g \right)^2 \approx tr (p-
{k\over 4\pi}\6 g \gi)^2$ and hence
 \be
   tr[p\6 g\gi ] \approx 0.
 \eel lemm5
This permits to eliminate the Hamiltonian in \rz fofac in agreement
with the fact that the model \rz agwzw is topological.

Being a Hamiltonian formulation of the GWZW model,  the form
\rz fofac
is not quite satisfactory, if one wants to solve the Gauss law
equation \re{const}). We therefore  apply here some
trick usually referred to as bosonization
\cite{GK,AS}.
The main idea is to substitute
the Gauss decomposition for the matrix $g$ into the GWZW action:
  \be
    g=\gdi \gu \ ,
 \eel decom
where $\gd$ is lower triangular, $\gu$ is
upper triangular, and both of them are elements of the
complexification of $G$. (Note, however, that we do not
complexify the target space $G$ here, but only use complex coordinates
$\gu$, $\gd$ on it). If the diagonal parts of $\gd$ and
$\gu$ are taken to be inverse to each other,
this splitting is unique up to sign ambiguities
in the evaluation of square roots.
Analogously any element of the Lie algebra $\cal G$ corresponding to
$G$
may be split into upper and lower triangular parts according to
 \be
   Y=Y\da .+Y\ua .\, ,\quad \left(Y\dA\right)_d =\left(Y\uA\right)_d =
\frac{1}{2} Y_d \
{},
 \eel algde
where a subscript $d$ is  used to denote the diagonal parts of the
corresponding matrices.

Observe that the three-form $tr (dgg^{-1})^3$  may be rewritten in
terms of $\gu$ and $\gd$ as follows:
 \be
\omega= {k\0 12\pi}  tr[(dg\gi )^3] =d \left[{k\0 4\pi} tr(d \gd \gdi\wedge
d\gu \gui)
\right]+ \varpi.
 \eel wzter
Here $\varpi$ is a three-form on $G$ supported at the lower dimensional
subset of $G$
which does not admit the Gauss decomposition. Now we can rewrite
the topological Wess-Zumino term as
\be
WZ(g)={k\over 12\pi}tr \int d^{-1}(dgg^{-1})^3=
{k\over 4\pi}tr \int d \gd \gdi\wedge d\gu \gui +
{k\over 12\pi}\int d^{-1}\varpi .
\eel WZvar
In this way we removed the symbol $d^{-1}$ in the first term
of the right hand side.
The  topological term
\be
S_\d={k\over 12\pi}\int d^{-1} \varpi
\eel var
is considered in details in Appendix B.
 In contrast with the conventional WZ term the new
topological term \rz var influences the equations of motion
only on some lower dimensional subset of the target space.

Let us return to the action of the GWZW model. We make the substitution
\rz decom and introduce a new momentum variable
 \be
    \Pi = \Pi\uA + \Pi\dA =
    \gd p  \gdi - {k\over 4\pi} \left(\6 \gu \gui + \6 \gd \gdi \right) \, .
 \eel defpi
Rescaling  $a_-$ according to $\lambda := {k \0 2\pi} a_-$,
we now may rewrite the GWZW action in the form
 \be \label{finac}
 GWZW(g, \Pi, \l)=S_{{\cal P}}(g, \Pi, \l)+ S_\d(g),
\end{equation}
where $S_\d$ has been introduced in (\ref{var}) and $S_{{\cal P}}$
is given by
 \begin{eqnarray}  \begin{array}{r@{}l}
 S_{{\cal P}}(g, \Pi, \l) &=\int d^2x\,
          tr\left\{\Pi\left(\6_0\gu\gui
              - \6_0\gd \gdi \right) +
         \right. \\[2pt] & \left.
            +\lambda\left[\gui \6_1\gu -
            \gdi\6_1\gd  +
            {2\pi \0 k} \left(\gui \Pi \gu -
               \gdi \Pi \gd \right)\right]\right\} \, . \label{finac2}
  \end{array}
 \end{eqnarray}
In the further consideration we systematically disregard
the topological term $S_\d$.
In Appendix C we prove that if we take \rz var into account,
the results change only for wave functions
 having support on those adjoint orbits in $G$ (one in the case of $G=SU(2)$)
 on which the Gauss decomposition breaks down.

For the formulation of a quantum theory in the $g$-representation, the
momentum $\Pi$ should be replaced by some derivative operator on the group.
The first term in \rz finac2 represents the symplectic potential on
the phase space and suggests the ansatz
   \be
g\to g \quad ,\qquad
\Pi \to - i (\gu {\d\over\d\gu} - \gd {\d\over\d\gd})
\, \, .   \eel qop

At this point  some remark on the notational convention is in
order: On $GL(N)$ coordinates are given by the entries $g_{ij}$ of the
matrix representing an element $g \in GL(N)$. The corresponding basis
in the tangent space may be arranged into matrix form via
   \be
\left(\frac{\delta}{\delta g} \right)_{ij} \equiv
\frac{\delta}{\delta (g_{ji})} \,\; .
   \ee
With this convention the entries of $g{\6 \0 \6 g}$ are
seen to be the right translation invariant vector fields on $GL(N)$.
Given a subgroup $G$ of $GL(N)$ the trace can be used to project the
translation invariant derivatives from $GL(N)$ to $G$. In more explicit
terms, given an element $Y$ of the Lie algebra of $G$, a right
translation invariant derivative on $G$ is defined by
$tr\,Yg{\6 \over\6 g}$. The matrix valued derivatives in this paper
are to be understood in this sense. In particular, \rz qop means that
the quantum operator associated to $tr\,Y\Pi$ is given by
   \be
tr\,Y{\Pi} \to -i                  \;
tr\; \left(Y\uA\gu {\d\over\d\gu} - Y\dA\gd {\d\over\d\gd}\right) \; .
   \ee
With this interpretation it is straightforward to prove that
commutators of
the quantum operators defined in \rz qop reproduce the Poisson algebra
of the corresponding classical observables, as defined by the
symplectic potential term in \ry finac2 .

Let us look for the wave functionals of the GWZW model in the
$g$-representation. This means that we must solve the equation
\begin{eqnarray} \label{gl}
(\gui \6_1\gu - \gdi\6_1\gd)\Psi(\gu, \gd)=  \qquad \qquad \qquad
\qquad \qquad
 \nonumber \\
       \frac{2\pi i}{k}\left(\gdi(\gd\frac{\delta}{\delta \gd}
- \gu\frac{\delta}{\delta \gu})\gd-
\gui (\gd\frac{\delta}{\delta \gd} -\gu\frac{\delta}{\delta \gu})\gu
\right) \Psi(\gu, \gd)
\end{eqnarray}
for $\Psi$ being a wave functional; the functional derivatives are
understood to act on $\Psi$ only (but not on everything to their right).
The problem is clearly
formulated, but at  first sight it is not evident how to solve
equation \re{gl}).  To simplify it we introduce another
parameterization of the matrix $g$:
\be
g= h^{-1}g_0 h.
\eel diag
Here  $g_0$ is diagonal and $h$ is defined up to an arbitrary
diagonal matrix  which may be multiplied from the left.  The part of the
operator \rz gl which includes functional derivatives simplifies
dramatically in terms of $h$. One can rewrite equation \rz gl as
\be
\left(\gui \6_1\gu - \gdi\6_1\gd + \frac{2\pi i}{ k } \frac{\delta}{\delta
h} h\right) \Psi[g_0, h] =0,
\eel hgl
where  $\gu, \gd$ are determined  implicitly as functions of $h$ and $g_0$
via \be \gdi \gu = h^{-1}g_0 h \, . \label{g} \ee

We discuss the interpretation of  equations \rz gl  and \rz hgl in
Section 2 and solve them efficiently in Section 3.

\section{Gauged WZW as a Poisson $\sigma$-Model}
The Gauss law equations of the previous section may be naturally
acquired in the theory of Poisson $\sigma$-models.
We start with a short description of this type of
topological $\sigma$-model.

The name Poisson $\sigma$-model originates from the fact that its
target space $N$ is a Poisson manifold, i.e.\ $N$ carries a
Poisson structure ${\cal P}$. We denote
coordinates on the two-dimensional world-sheet $M$ by $x^{\mu},
\mu=1,2$
and coordinates on the target space $N$ by $X^i, i=1,\dots ,n$. A
Poisson bracket $\{ \, \cdot \,  , \,  \cdot \, \}$
on $N$ is defined by specifying its value
for some coordinate functions: $\{X^i,X^j\} =
{\cal P}^{ij}(X)$. Equivalently the
Poisson structure may be represented by a bivector
\be
{\cal P}=\2 {\cal P}^{ij}(X) \frac{\partial}{\partial X^i}\wedge
\frac{\partial}{\partial X^j} \, .
\eel Poiss
In terms of this tensor the  Jacobi identity for $\{ \, \cdot \,  ,
\,  \cdot \, \}$ becomes
\be
{\cal P}^{li}\frac{\partial {\cal P}^{jk}}{\partial X^l} +
{\cal P}^{lk}\frac{\partial {\cal P}^{ij}}{\partial X^l} +
{\cal P}^{lj}\frac{\partial {\cal P}^{ki}}{\partial X^l} =0 \ .
\eel Jid
For nondegenerate
 ${\cal P}$ the notion of a Poisson manifold coincides with that of
 a symplectic manifold. In general,  however,
${\cal P}$ need not be nondegenerate.

In the world-sheet picture our dynamical variables are the $X^i$'s and a
field $A$ which is a one-form in both world-sheet and target space. In
local coordinates $A$ may be represented as
\be
A=A_{i \mu} dX^i \wedge dx^{\mu}.
\eel defA
The topological action of the Poisson $\sigma$-model consists of two
terms, which we write in coordinates:
\be
S_{{\cal P}}(X, A)=\int_{M}\left( A_{i \nu} \frac{\partial
X^i}{\partial x^\mu}  +
\frac{1}{2} {\cal P}^{ij} A_{i \mu} A_{j \nu}\right) dx^{\mu} \wedge
dx^{\nu} \, .
\eel defS
Here $A$ and $X$ are understood as functions on the world-sheet.
Both terms in \rz defS are two-forms with
respect to the world-sheet. Thus,
they may be integrated over $M$. The action \rz defS is obviously
invariant with respect to diffeomorphisms of the world-sheet.
It is also invariant under diffeomorphisms of the target space
which preserve the Poisson tensor.
Equations of motion for the fields $X$ and $A$ are
\begin{eqnarray}
\partial_{\mu} X^i+{\cal P}^{ij}A_{j\mu}=0 \, ,
\nonumber \\
\partial_{\mu} A_{i\nu}-\partial_{\nu} A_{i\mu} -
\frac{\partial {\cal P}^{jk}}{\partial X^i}A_{j\mu} A_{k\nu}=0 \, .
\end{eqnarray}
Here $\partial_{\mu}$ is the derivative with respect to $x^{\mu}$ on the
world-sheet.

Let us remark that the two-dimensional BF theory may be interpreted
as a Poisson $\sigma$-model. Indeed, if one chooses a Lie algebra
with structure constants $f^{ij}{}_k$ as the
target space $N$ and
uses the natural Poisson bracket
\be
\{ X^i , X^j\}=f^{ij}{}_k X^k ,
\eel Liealg
one reproduces the action
of the BF theory
\be
BF(X, A)=\int_M tr X(dA+A^2).
\eel BF
In the traditional notation $X$ is replaced by $B$ and the
curvature $dA+A^2$ of the gauge field is denoted by $F$.
The class of Poisson $\sigma$-models includes also nontrivial
examples of two-dimensional theories of gravity (for details see
\cite{Th2,Th}).

We argue that the gauged WZW model is equivalent to a Poisson
$\sigma$-model coupled to the term (\ref{var}).
The target space is the group $G$, parameterized by
$\gu$ and $\gd$.
The $(1, 1)$-form $A$ is identified readily from \re{finac2}):
  \be
  A = \Pi \left( d\gu \gui  -  d\gd \gdi \right)\wedge dx^1
   -\l  \left( \gui d\gu  -  \gdi d\gd \right)\wedge dx^0 \, .
\label{A} \ee
Here we have interpreted the terms linear in $\Pi$ and $\l$.

Then the part of the action quadratic in $\Pi$ and $\l$ directly
determines the Poisson structure.
In our formulation of the general
Poisson $\sigma$-model \rz defS the indices $i,\m$ of $A$ correspond to a
coordinate basis $dX^i$ in $T^*N$ and
$dx^\m$ in $T^* M$.
In such a formulation we simply have to replace
$A_{i\mu}$ by ${\6\over\6X^i}$ in the quadratic part of the action
to obtain the Poisson bivector \rz Poiss as the 'coefficient' of the
volume-form $dx^\m \wedge dx^\n$. Each of the matrix-valued one-forms
$ d\gu \gui  -  d\gd \gdi $ and $\gui d\gu  -  \gdi d\gd $ in
the present expression \rz A for $A$, however, represents a
non-holonomic basis in the cotangent bundle of the target space $G$.
In such a case the corresponding components of $A$, i.e.\
 $\Pi$ and $\l$ in our notation,  have to be replaced by the
respective dual derivative matrices.  Applying this
simple recipe to  the quadratic part of \ry finac2 , we find the
Poisson bivector on $G$:
   \be
\begin{array}{c}
 \Pi \to (\gu {\6\over\6\gu} - \gd {\6\over\6\gd})\, ; \quad
 \l \to ({\6\over\6\gu} \gu -  {\6\over\6\gd} \gd) \\[5pt]
 \Rightarrow \quad
{\cal P} = {4\pi\over  k }\,
tr \, \left( {\6\over\6\gu} \gu - {\6\over\6\gd} \gd
 \right) \wedge \\[2pt]
\left(\gui (\gd \frac{\partial}{\partial \gd} -\gu
\frac{\partial}{\partial \gu} )\gu - \gdi (\gd
\frac{\partial}{\partial \gd} -\gu \frac{\partial}{\partial \gu})\gd
\right) \  .
\end{array}
\eel WZWPois
Using the parameterization
 (\ref{diag}, \ref{g}),  this expression can be
formally simplified to
\be
  {\cal P} = {4\pi \over  k }\, tr \,
  \left( {\6\over\6\gu} \gu - {\6\over\6\gd}
  \gd \right) \wedge {\6\over\6 h} h \ .
\eel WZWPois2
For means of completeness we should  check now
that this bivector fulfills
the Jacobi identity \ry Jid . In our context  the simplest way to do so is
to recall that the constraints of the GWZW model are first class;
this suffices, because one can show that the constraints of any
action of the form \rz defS are first class exactly
iff ${\cal P}^{ij}$ obeys \ry Jid . Certainly
one can verify the Jacobi identity also
by some direct calculation and in fact this is done implicitly when
establishing \rz clYB and \rz G* below.

The above Poisson bracket on $G$ requires  further
comment. For this purpose it is useful to introduce some new object.
We always assume that  the group $G$ is realized as a subgroup in the
group of $n$ by $n$ matrices. Then the following matrix $r$ acting in
$C^n\otimes C^n$ is important for us:
\be
r=\frac{1}{2}\sum_i h^i\otimes h_i +
   \sum_\alpha t_{-\alpha}\otimes t^\alpha.
\eel rmatrix
Here $h^i$ and $h_i$ are generators of dual bases in the Cartan
subalgebra, $t^\alpha$ and $t_{-\alpha}$ are positive and negative
roots, respectively. The matrix $r$ is usually called classical
$r$-matrix. It satisfies the classical Yang-Baxter equation in the
triple tensor product which reads
\be
[r_{12} , r_{13}] + [r_{12} , r_{23}] + [r_{13} , r_{23}]=0.
\eel clYB
Here $r_{12}=r \otimes 1$ is embedded in the product of the first two
spaces and so on.  An important property of the $r$-matrix is the
following
\be
tr_{1,2} r A^1 B^2= tr \, A_\uparrow B_\downarrow + \frac{1}{2} tr \, A_d
B_d \, ,
\eel Traces
where the trace in the left hand side is evaluated in the tensor product of
two spaces and
\be
A^1 \equiv A\otimes 1 \ , \ B^2 \equiv 1\otimes B \ .
\eel tens12

Now we are ready to represent the bracket \rz WZWPois in a more
manageable way.
As the most natural coordinates on the group are
matrix elements, we are interested in Poisson brackets of  entries of
$\gu$ and $\gd$.  Using shorthand notations \rz tens12 and the
definition of the $r$-matrix,  we  arrive at the following elegant
result
\begin{eqnarray}  \label{G*}
\{ \gu^1 , \gu^2\}=\frac{4\pi}{k}[r, \gu^1 \gu^2] \ , \nonumber \\
\{ \gd^1 , \gd^2\}=\frac{4\pi}{k}[r, \gd^1 \gd^2] \ ,  \\
\{ \gd^1 , \gu^2\}=\frac{4\pi}{k}[r, \gd^1 \gu^2] \ . \nonumber
\end{eqnarray}
We omit the calculation which leads to  \ry G* , as it is  rather
lengthy but straightforward. Each equation in \rz G* provides
a Poisson bracket between any matrix element of the matrix with superscript
$1$ with any matrix element of the matrix with superscript $2$.
In order to clarify this statement, we rewrite the first equation
in components:
\be
\{ \gu^{ij} , \gu^{kl}\}={4\pi \over  k }
(r^{i\tilde{i}}_{k\tilde{k}}\gu^{\tilde{i}j}
\gu^{\tilde{k}l}-\gu^{i\tilde{j}} \gu^{k\tilde{l}}r^{\tilde{j}j}_{\tilde{l}l}).
\eel r-comp
Here summation over the indices with tilde in the right hand side is
understood.
The remaining equations of \rz G* can be rewritten in the same fashion.
The formulae \rz G* define the Poisson bracket only on the subset
of the group $G$ which admits the Gauss decomposition \ry decom .
One can easily recover the Poisson brackets of matrix
elements of the original matrix $g$. We leave this as an exercise
to the reader. The result may be presented in tensor notation:
\be
\{ g^1, g^2\}=\frac{4\pi}{ k }[g^1 r g^2 + g^2 r' g^1 - r' g^1 g^2 -
g^1 g^2 r].
\eel 4r
Here $r'$ is obtained from $r$ by exchanging the two copies of the
Lie algebra:
\be
r'= \frac{1}{2}\sum_i h^i\otimes h_i +
   \sum_\alpha t_\alpha \otimes t^{-\alpha}.
\eel r'
The Poisson bracket \rz 4r is quadratic in matrix elements of $g$
and obviously smooth. This means that the bracket \rz G* which
has been defined only on the part  of the
group $G$ where the Gauss decomposition is applicable,  may
now be continued smoothly to the whole group.
E.g.\ for the case of $G=SU(2)$ it is straightforward to establish
that the right-hand side of \ry 4r , and thus also the smoothly continued
Poisson tensor ${\cal P}$,   vanishes at antidiagonal matrices $g \in SU(2)$.
The latter represent precisely
 the one-dimensional submanifold of $SU(2)$ where a decomposition
(\ref{decom}) for $g$  does not exist.
It is worth mentioning
that the bracket \rz 4r appeared first in \cite{Sem}
within the framework of the theory of Poisson-Lie groups.

The group $G$ equipped with the Poisson bracket \rz 4r may be used
as a target space of the Poisson $\sigma$-model. We have just proved
that in the Hamiltonian formulation (geometry of the world-sheet is
torus or cylinder) this Poisson $\sigma$-model  coupled
to the topological term \rz var  coincides with
the gauged WZW model.

\section{Solving the Gauss Law Equation}
This section is devoted to the  quantization of Poisson $\sigma$-models.
More exactly, we are interested in the Hilbert space of such a model
in the Hamiltonian picture. This implies that we need a distinguished
time direction on the world-sheet and thus we are dealing with a
cylinder. The remarkable property of Poisson $\sigma$-models
is that the problem of finding the Hilbert space in this
two-dimensional field theory may be actually reduced to a quantum
mechanical problem.  This has been realized in \cite{Th} and here we
give only a short account of the argument\footnote{But cf.\ also
Appendix C.}.

It follows form \rz defS that in the  Hamiltonian formulation
the variables
$X^i$ and $A_{i1}$ are canonically conjugate to each other
(this changes slightly when the Poisson $\sigma$-model
is coupled to the term $S_\d$, see Appendix C). In the
$X$-representation of the quantum theory  the $X^i$ act as
multiplicative operators and the
$A_{i1}$ act
as functional derivatives
\be
A_{i1}=i\frac{\delta}{\delta X^i} \ .
\eel A=dX

The components  $A_{i0}$ enter the action linearly. They are
naturally interpreted as  Lagrange multipliers. The corresponding
constraints look as
\be
G^i \equiv \partial_1 X^i + {\cal P}^{ij}(X) A_{j1} \, \approx \, 0 \; .
\eel consXA
Combining \rz A=dX and \ry consXA , one
obtains an equation for the wave
functional in the $X$-representation
\be
\left(\partial_1 X^i + i {\cal P}^{ij}(X)\frac{\d}{\d
X^j}\right)\Psi[X]=0 \ .
\eel PsiX
Equations \rz gl and \rz hgl are particular cases of this equation.
In order to solve \ry PsiX , we first turn to
a family of finite dimensional
systems on the target space $N$ defined by the Poisson structure ${\cal P}$.

As the target space of a Poisson $\sigma$-model carries a Poisson
bracket, it may be considered as the starting point of a
quantization problem. Namely, one can consider the target space as
the phase space of a finite dimensional Hamiltonian system, which one
may try to quantize. The main obstruction on this way is that the
Poisson bracket ${\cal P}$ may be degenerate. This means that if
we select some point in the target space and then move it by means
of all possible Hamiltonians, we still do not cover the whole
target space with trajectories but rather stay on some surface
${\cal S} \subset N$.  The simplest example of such a situation is
a three-dimensional space $N=\dR^3$ with the Poisson bracket \be
\{ X^i , X^j \}=\epsilon_{ijk} X^k \ .  \eel su2 This Poisson
bracket describes a three-dimensional angular momentum and it is
well-known that the square of the length \be R^2:=\sum_i (X^i)^2
\ee commutes with each of the $X^i$. So, $R^2$ cannot be changed
by means of Hamiltonian flows and the surfaces ${\cal S}$ are
two-dimensional spheres.

If the Poisson bracket ${\cal P}$ is degenerate, we cannot use $N$ as
a phase space. However, if we restrict to some surface
${\cal S}$ (these surfaces are also called symplectic leaves),
the Poisson bracket becomes nondegenerate and one can try to
carry out the quantization program. In the functional integral
approach we are interested in the exponent $\exp(i{\cal A})$ of the
classical action ${\cal A}$,
being the main ingredient of the quantization scheme. In
order to construct the classical action ${\cal A}$, we invert the matrix of
Poisson brackets (restricted to some particular surface ${\cal S}$)
and obtain a symplectic two-form
\begin{eqnarray} \label{sympl}
\O=\frac{1}{2} \O_{ij} dX^i\wedge dX^j \ , \nonumber \\
\sum_k \O_{ik} {\cal P}^{jk}=\delta_i^j \ .
\end{eqnarray}
As a consequence of the Jacobi identity the form $\O$ is closed
\be
d \O=0 \
\eel closed
and we can look for a one-form $\a$ which solves the equation
\be
d\a=\O \ .
\eel pdq
If $\O$ belongs to some nontrivial cohomology class, $\a \sim pdq$
does not exist globally. Still the
expression
\be
\Psi[X]:=exp\left(i \int d^{-1}\O \right)
\eel ansatz
makes sense,
if the cohomology class of $\O$ is integral, i.e. if
\be
\oint_\sigma \O = 2\pi n \; , \quad n \in Z
\eel 63
for all two-cycles $\sigma \subset
{\cal S}$;
in this case ${\cal A}=\int d^{-1}\O$ is defined
$\mbox{mod} \; 2\pi$ and \rz ansatz is
one-valued (cf.\ also \rz defA(X) below).
Alternatively to the functional integral approach
we might use the machinery of geometric quantization \cite{Wood}
to obtain condition \ry 63 : Within the approach of geometric quantization
it is a well-known fact that a Hamiltonian system $({\cal S}, \O)$ may be
quantized consistently only if the symplectic form $\O$ belongs to
an integral cohomology class of ${\cal S}$.
In the example of two-dimensional spheres in the three-dimensional
target space
considered above the requirement of the
symplectic leaf to be quantizable, obtained in any of the two approaches
suggested above,  implies that the  radius $R$ of the sphere is either
integer or half-integer
(for more  details confer \cite{NR,Wood}).
This  is a manifestation
of the elementary  fact that a three-dimensional spin has to be either
integer or half-integer.

After this excursion into  Hamiltonian mechanics we return
to equation \re{PsiX}).
It is possible to show that
formula \rz ansatz provides a solution of equation
\ry PsiX . Moreover, any solution of
\rz PsiX  can be represented as a linear
combination of  expressions \rz ansatz corresponding
to different integral symplectic leaves \cite{Th}.

Let us explain this  in more detail.
The wave functional $\Psi[X]$ of the field theory depends on $n$ functions
$X^i$ on the circle.
They define a
parameterized closed trajectory (loop) in the target space $N$. Now
it is a more or less immediate consequence of \rz PsiX
 that the quantum constraints of the field theory restrict the support of
$\Psi$ to  trajectories (loops) $X(x^1)$ which lie completely within
a symplectic leaf ${\cal S}$ (just use coordinates $X^i$ in the
target space adapted to the foliation of $N$ into symplectic leaves).
A further analysis, recapitulated in part in Appendix C within the
more general framework of a Poisson $\s$-model coupled to a
topological term, shows that
these leaves have to be  quantizable  and that
admissible quantum states are indeed all of the form \rz ansatz or a
superposition of such functionals. In the case that ${\cal S}$ is
simply connected, \rz ansatz may be rewritten more explicitly as:
\be
\Psi[X] \propto \exp (i{\cal A}(X)) \;  \; , \;\; \quad {\cal A}(X) =
\int_\Sigma \O  \quad (\mbox{mod} \; 2 \pi) \; ,
\eel defA(X)
where the two-dimensional surface $\S$ is bounded by the closed path
$X(x^1)$ lying in some quantizable leaf ${\cal S}$.\footnote{In the
language of Appendix C the definition \rz defA(X)
corresponds to the choice of a constant (point-like) 'loop of
reference' for $\Psi_0$.}
As $\O$ belongs to an integral cohomology class (by the choice of
${\cal S}$),
\rz defA(X) is a  globally well-defined  functional of $X(x^1)$. As stated
already before any such a functional solves the quantum constraints
\re{PsiX}) and, vice versa, any solution to the latter has to be a
superposition of states \ry defA(X) .
On the other hand  \rz ansatz or \rz defA(X) may be also reinterpreted
as exponentiated point particle action. $x^1$ then is the 'time-parameter'
of the trajectory $X(x^1)$, which one requires to be periodic in time.

So we  obtain the following picture for
the relation between the Poisson $\sigma$-model
and finite dimensional quantum mechanics:
In order to
obtain the Hilbert space of the $\s$-model
on the cylinder, one may
regard  the target space as a phase space of a dynamical system.
This space splits into a set of surfaces on which the
Poisson bracket is nondegenerate, creating a family of finite
dimensional systems. Some of these systems are quantizable in the sense that
the cohomology class of the symplectic form is integral.
To each quantum system generated in this way
corresponds a linearly independent vector in the Hilbert space
${\cal H}$ of the
$\s$-model. In the case that the respective
(quantizable) symplectic leaf ${\cal S}$ is not simply connected,
however,
there is a linearly independent vector in ${\cal H}$
 for {\em any} element of $\pi_1({\cal S})$. This idea may be successfully
checked for BF theories
in two dimensions (for more details confer \cite{Th}).

Now we apply the machinery of this section to the GWZW model. First, we
should look at  the surfaces ${\cal S}$ in the group $G$ where the
restriction of the Poisson bracket \rz G* is nondegenerate.
For generic leaves this problem has been
solved in \cite{Sem}. In order to make ${\cal P}$ nondegenerate, one
should restrict to some conjugacy class in the group
\be
g=h^{-1}g_0h \ .
\eel conju
Each conjugacy class may be used  as the phase space of a Hamiltonian
system. However, in the case of $G=SU(2)$ we found
that the Poisson bracket vanishes on the subset  of
antidiagonal matrices. Hence, any antidiagonal matrix represents a
zero-dimensional symplectic leaf in $G=SU(2)$. So, some exceptional
conjugacy classes may further split into families of symplectic leaves.
This occurs precisely where the Gauss decomposition does not hold.

The form $\O$ on a generic  orbit characterized by $g_0$
has been recently evaluated in \cite{AM} (a presentation
more adapted to the physical audience can be found in \cite{AT}) and  has
the form
\be
 \Omega =
{ k \over 4\pi} tr\left[h^{-1}dh\wedge (\gui d\gu -\gdi d\gd)\right] \, ,
\eel omega
where $\gu$, $\gd$, and $h$ are related through \ry g .
The corresponding point particle action or phase factor of the
quantum states, respectively, is
\be
{\cal A}_{GWZW}(g)={ k \over 4\pi}
\int d^{-1} tr\left[h^{-1}dh\wedge (\gui d\gu
- \gdi d\gd)\right] \ .
\eel geom

As outlined above quantum states are assigned only to integral
symplectic leaves. In Appendix C  the corresponding
integrality condition \rz 63  is evaluated explicitly for the example of
$G= SU(2)$.

The exceptional conjugacy classes require some special
attention. From the point of view of the {\em pure}
Poisson $\sigma$-model
there corresponds a quantum state to any integral symplectic leaf
which the respective conjugacy class may contain. For the case of
$SU(2)$, e.g., there is one exceptional (two-dimensional)
conjugacy class \rz conju characterized by \mbox{$tr \, g=0$}.
It contains the one-dimensional
submanifold ${\cal C}$ of antidiagonal matrices in $SU(2)$. Any
point of ${\cal C}$ is a zero-dimensional symplectic leaf and,
because zero-dimensional leaves are always quantizable, one would
be left with a whole bunch of states corresponding to this
exceptional conjugacy class.

However, we know that in order to describe the GWZW model
in full generality, we need to add the topological term $S_\d$ to the pure
Poisson $\sigma$-part of the action. Also, appropriate boundary conditions
of $A$ have to
be taken into account at the part of $G$ where the Gauss decomposition
breaks down. Whereas $S_\d$ and these boundary conditions may be seen to
be irrelevant for the quantum states corresponding to generic
conjugacy classes, they decisively change the picture at the
exceptional ones. E.g.\ for $G=SU(2)$ the net result is that
there corresponds only  one or even  no
quantum state to the exceptional conjugacy class, depending on
whether  $k$ is even or odd, respectively.
Further details on this may be found in  Appendix C.

{}From \rz geom
it is straightforward to evaluate the cocycle $\phi_{GWZW}$
which controls the behavior of the wave functional with respect
to gauge transformations. E.g., for the case of infinitesimal
transformations
\be
\delta g= - [\epsilon, g] \ , \ \delta h=  h\epsilon
\eel infgauge
the new gauge cocycle looks as:
\be
\phi_{GWZW}(g, \epsilon)={ k \over 4\pi} \int
 tr \,  \epsilon \: (\gui d\gu -\gdi d\gd)\ .
\eel WLu
An integrand  of this type has been studied in the framework of
Poisson-Lie group theory \cite{WLu}. However, the gauge
algebra interpretation is new.

In order to check that the cocycle $\phi_{GWZW}$ is nontrivial,
it is
convenient to use the same trick as we applied in Introduction.
Indeed, choose both $g$ and $\epsilon$ to
be diagonal. Then any trivial cocycle vanishes, but \rz WLu
is  not equal to zero for generic diagonal $g$ and~$\epsilon$.

\section*{Discussion}
Let us briefly recollect and discuss the results of the paper.
Using the Hamiltonian formulation we have proved that the GWZW
model is equivalent to a Poisson $\sigma$-model coupled to the
topological term $S_\d$: \be GWZW(g, A)=S_{\cal P}(g, A)+S_\d(g).
\eel S+S2 It is natural to conjecture that this equivalence holds
true for a surface of arbitrary genus.  Let us mention that
originally the GWZW is formulated as a Witten type topological
field theory. This means that the action includes the kinetic term
and explicitly depends on the world-sheet metric. Then one can use
some supersymmetry to prove that in fact the terms including the
world-sheet metric do not influence physical correlators. The
Poisson $\sigma$-model provides a Schwarz type formulation of the
same theory. The right hand side of \rz S+S2 is expressed in terms
of differential forms exclusively and does not include any metric
from the very beginning.

At the moment the GWZW model is solved in many ways
whereas the general Poisson $\sigma$-model has not
been investigated much. Applying various methods
which work for the GWZW to Poisson $\sigma$-models,
one can hope to achieve two goals. First, one can select the
methods which work in a more general framework and, hence, which
are more reliable. This is especially important when one
deals with functional integrals. The other ambitious program
is to solve an arbitrary Poisson $\sigma$-model coupled
to a topological term explicitly. Solution should include an
evaluation of the partition function and topological
correlators in terms of the data of the target space.
In this respect an experience of the GWZW model may be
very useful.

Another issue which deserves some comment is the relation
between Quantum Groups and WZW models. This issue has
been much studied in the literature \cite{BB}.
The picture of the quantum symmetry
in WZW models may be described in short as follows.
Separating left-moving and right-moving sectors of the
model we add some finite number of degrees of
freedom to the system. The Quantum Group symmetry
is a gauge symmetry acting on the left- and right-movers.
The physical fields are invariants of the Quantum Group
action. Usually one can choose some special boundary
conditions when separating the sectors in order to
make the Quantum Group symmetry transparent.

Let us compare this picture to the considerations of the present paper.
The gauged WZW model appears to be equivalent to some
Poisson $\sigma$-model with gauge group $G$ as  target
space. We derive the Poisson bracket \rz 4r directly
from the GWZW action. This bracket is quite remarkable.
Quantizing the bracket \ry 4r , one gets the generating
relations of the Quantum Group \cite{FRT}. We have
found that the gauge dependence of the wave functionals
of the GWZW model is described by the classical action
defined on the symplectic leaves. This type of action
for the bracket \rz 4r has been considered in \cite{AT}.
It is proved there that such an action possesses a symmetry
with respect to the Quantum Group. So, confirming our
expectations, the Quantum Group governs the non-physical
gauge degrees of freedom of the GWZW model. The new
element of the picture is that we do not have to introduce
any new variables or choose specific boundary conditions
in order to discover the Quantum Group structure.
Let us remark that the treatment may look somewhat
more natural for GWZW than for the  original WZW model.
The reason is that GWZW may be viewed as a chiral theory
from the very
beginning\footnote{We are grateful to K.Gawedzki for this remark.}.
The only
choice which we make is  the way how we bosonize the WZW
action. We conclude that the Quantum Group degrees of freedom
are introduced by bosonization. It would be interesting to
explore this idea from a more mathematical point of view.

\newpage
\section*{Appendices}

\begin{appendix}
\renewcommand{\theequation}{\thesection.\arabic{equation}}
\setcounter{equation}{0}
\section{Gauge Cocycles and Integral Coadjoint Orbits}
Here we study in details the one-cocycle
\be
\phi(E, g)=\oint tr \,  E \partial gg^{-1} dx
\eel Aphi
of the loop group $LG$, which plays the role of the
gauge group on the circle. Along with the additive
cocycle $\phi$ we consider a multiplicative cocycle
\be
\Phi(E, g)=\exp (i \phi(E, g)) \, .
\ee
The counterparts of the cocycle and coboundary conditions
in the multiplicative setting are
\ba
\Phi(E,g_1g_2)&=&\Phi(E^{g_1}, g_2) \Phi(E,g_1) \, , \\
\Phi(E, g)&=&\tilde{\Phi}(E^g)\tilde{\Phi}(E)^{-1} \, . \label{GL}
\ea

Let us observe that  one can consistently
restrict the region of definition of $E$ from the
loop algebra $l{\cal G}$ to any
subspace invariant with respect to the action of the gauge
group by conjugations. Let us choose such a subset
in the form
\be
E=h(x)^{-1} E_0 h(x)
\eel coorb
for $E_0$ being a  constant diagonal matrix. For  fixed $x$
equation \rz coorb defines a conjugacy class in the algebra
${\cal G}$ (coadjoint orbit).

The diagonal matrix $E_0$ may be decomposed using a basis
of fundamental weights $w_i$ in the Cartan subalgebra:
\be
E_0= \sum_i E_0^i w_i.
\eel int
In the case of compact groups the cocycle $\Phi$ is trivial if and only
if all coefficients $E_0^i$ are integer. To demonstrate this, let
us present the explicit solution for
$\tilde{\Phi}$. It is given by
\be
\tilde{\Phi}=\exp \left(i \oint  tr \, E_0 \partial hh^{-1} dx \right) \,.
\eel Atil
It is easy to check that \rz Atil  provides a
solution of the coboundary problem. It is less evident that
\rz Atil  is well-defined. The group element $h(x)$ is defined
by $E(x)$ only up to an arbitrary diagonal left multiplier.
When coefficients in \rz int are integral,
this multiplier does not influence \ry Atil .

For non-compact groups, though,
\rz Atil may turn out to be well-defined even for continuously varying
choices of $E_0$.

To establish contact with
the presentation in the main text, one may observe that
the additive coboundary
(generically not well-defined)
\be
\tilde{\phi}=\oint tr \, E_0 \partial hh^{-1}  dx
\eel plops
may be reformulated in terms of the (well-defined)
Kirillov form on the coadjoint orbit,
\be \O =      tr \, E_0 (dhh^{-1})^2 =  \2 \, tr \, dE \wedge h^{-1}dh
\, ,  \eel Kir
as
\be \ti \phi = \int_\S\O \; , \quad \6 \S = E(x) \, .
\eel formel
The ambiguity in the choice of $\Sigma$ does not influence the
multiplicative  cocycle
$\tilde\Phi$, iff the Kirillov form is integral, i.e.\ iff $\O$ satisfies
\rz 63 .

For compact groups
the integrality condition \rz 63 on $\O$ coincides with the
before-mentioned condition on the $E_0^i$.
If \rz 63 is fulfilled with $n = 0$ not only the multiplicative
but also the additive cocycle $\phi$ becomes trivial.
This occurs, e.g., in the non-compact case ${\cal G}=sl(2,\dR)$).

It is worth mentioning that \rz plops
is the action for a quantum mechanical
system with the phase space being a coadjoint orbit.
We consider a similar system in Section 3.  There the quantum
mechanical phase space is a conjugacy class in the group and the
analogue of the  Kirillov form \rz Kir is \ry omega ,
the Kirillov form of the Quantum Group.

We conclude that for certain restricted subspaces
of the loop algebra the cocycle $\Phi(E, g)$ may become trivial.
In two dimensions the wave functionals in the momentum representation
are supported on these special subspaces. The corresponding
coboundary $\tilde{\Phi}$ governs the gauge dependence
of the wave functionals:
\be
\Psi= \tilde{\Phi} \,   \Psi_0
\eel tPhi
for $\Psi_0$ being a gauge independent distribution with support on
loops in integral coadjoint orbits.

Let us stress again that the triviality condition
\rz GL is actually an integrated form of the Gauss law (as shown in the
introduction).
Then \rz tPhi provides a universal solution of the
Gauss law. In the example which we considered in this
Appendix we observe a new phenomenon in the theory
of gauge cocycles. A nontrivial cocycle may shrink
its support in order to become trivial and produce
a physical wave functional. This may lead (as in the example of 2D YM
theory with compact gauge group)
to a discrete spectrum in the momentum representation.

\section{Topological Term for $G=SU(2)$}
\setcounter{equation}{0}
The topological Wess-Zumino term in the WZW model is usually
represented in the form
\be
WZ(g)=\frac{ k }{12\pi} tr \int_{\Sigma} d^{-1}(dgg^{-1})^3.
\eel WZ
The integration is formally performed over the  two-dimensional
surface $\Sigma$. (Here $\Sigma$ is the image of the world sheet
$M$ under the map $g(x)$ from $M \ra G$).
The symbol $d^{-1}$ has been introduced by Novikov \cite{Nov}
and applied to construct miltivalued action functionals
in \cite{Nov}, \cite{JW}.
It is understood in the following
way. One chooses a three-dimensional submanifold $B$ in the group $G$
so that $\partial B=\Sigma$. The integration over $\Sigma$ is replaced
by an integration over $B$:
\be
WZ(g)=\frac{ k }{12\pi}tr \int_B (dgg^{-1})^3.
\eel WZB
The definition \rz WZB is ambiguous as $B$ may be chosen in many ways.
The possible ambiguity in the definition of $WZ(g)$ is an integral
over the union of two possible $B$'s:
\begin{eqnarray} \label{DWZ}
\Delta WZ(g)&=&\frac{ k }{12\pi}
tr (\int_{B'} (dgg^{-1})^3 - \int_{B''}
(dgg^{-1})^3) \; = \nonumber \\
&=& \frac{ k }{12\pi} tr \int_{B'\cup \bar{B''}} (dgg^{-1})^3 \; .
\end{eqnarray}
Here we denote by $\bar{B''}$ the manifold $B''$ with opposite
orientation. Let us restrict our
 consideration
to the case of $G=SU(2)$. The only
nontrivial three-dimensional cycle in $SU(2)$ is the group itself.
It implies that
the  integral (\ref{DWZ})  is always proportional
with some integer coefficient    to the normalization
integral
\be
I=\frac{ k }{12\pi}tr \int_G (dgg^{-1})^3= 2\pi  k  \; .
\eel WZnorm
Here we used the fact that the volume of the
group $SU(2)$ with respect to the form $tr(dgg^{-1})^3$ is equal to
$24\pi^2$. This calculation explains why one should choose integer
values of the coupling constant $ k $. In this case the Wess-Zumino
term $WZ(g)$ is defined modulo $2\pi$ and its exponent $\exp(iWZ(g))$
is well-defined.

Usually $WZ(g)$ is referred to as a topological term because the
defining three-form $tr (dgg^{-1})^3$ on the group $G$ is closed
and belongs to a nontrivial cohomology class. This implies that
the integral (\ref{WZ}) does not change when we fix $\Sigma$ and
vary $B$ in a smooth way. Choosing the proper coefficient
$k/12\pi$, $k \in N$,
 we get a three-form which belongs to an integer
cohomology class. As we have seen this ensures that $\exp(iWZ(g))$
is preserved even by a topologically nontrivial change of $B$.

So the fact that the three-form
$$ \omega=\frac{k}{12\pi}
tr (dgg^{-1})^3 $$
is closed
and belongs to integer cohomology of $G$ makes the action $WZ(g)$
well-defined. However, it is not true that $WZ(g)$ is defined
already by the cohomology  class of $\omega$.
If we choose some other representative
in the same class (as, e.g., $\varpi$ in Eq.\ \rz gov below),
we get a new topological term, which is well-defined
for the same reason as $WZ(g)$. In fact, the new action
will differ from $WZ(g)$. The reason  is that the
integral \rz WZB is defined over the manifold with a boundary and,
hence, it is not defined by the cohomology class of the integrand. It
depends on the representative as well.

Now we are prepared to introduce a new topological term for the WZW
model. As it was explained in Section 1, we use the Gauss decomposition
for the group element $g$:
\be
    g=\gdi \gu \ .
\eel decomA
Observe that the Gauss decomposition is not applicable for some elements
in $SU(2)$.
The Gauss components $\gd, \gu$ do not exist on
the subset  of
antidiagonal unitary matrices.
In a parameterization
  \be
     g=\left(\begin{array}{cc} z &\sqrt{1-z\bar z}\,e^{i\phi} \\[6pt]
                 -\sqrt{1-z\bar z}\,e^{-i\phi}  & \bar z
            \end{array}\right)\quad , \qquad
        z\in C,\; |z| \le 1,\; \phi \in [0,2\pi )
   \eel parm
these elements are given by $z=0$. They form a circle $\cal C$
parameterized by $\phi$.

We can apply the Gauss decomposition on the rest of
the group in order to remove the symbol $d^{-1}$ from the topological
term $\o$. Indeed, consider a two-form
\be
\gamma={k \0 4\pi} \, tr\, (d \gd \gdi \wedge d\gu \gui )
\eel gamma
on the compliment of $\cal C$. It is easy to verify  the relation
\be
d\gamma= \frac{1}{3} \omega.
\eel gamom
Here we have used the fact that
\be
tr (d \gd \gdi)^3= tr (d\gu \gu)^3=0,
\eel vanish3
which holds since the diagonal parts of $(dM M^{-1})^m$
vanish for any triangular matrix $M$ if $m \ge 2$.

We established equation \rz gamom on the part of the group $G$
which admits the Gauss decomposition. It is easy to see that
this equation cannot hold true on all of $G$.
Indeed, the left hand side is represented by the exact form
$d\gamma$ whereas the right hand side belongs to a nontrivial
cohomology class. In order to improve \ry gamom , we introduce
a correction to it:
\be
d\gamma=\frac{1}{3}(\omega- \varpi).
\eel gov
This equation is to be understood in a distributional sense:
The three-form $\varpi$ is  supported
on ${\cal C}$. Moreover it is closed and belongs to the same
cohomology class as $\omega$.

To determine $\varpi$ for  $G=SU(2)$, we return to
the parameterization \ry parm . In these coordinates  \rz gamma
takes the form
\be \g =  i \left(\bar z  dz - z d \bar z -2 {dz \0 z} \right)
d \phi  \; . \ee
Multiplying $\g$ by test one-forms, the resulting three-forms
are integrable on $G$. So $\g$ is a regular distribution and
therefore  the  derivative $d\g$ is also
well-defined.  Using
$d ( dz/z)= \pi \d(\mbox{Re}(z))\d(\mbox{Im}(z)) dzd \bar z =:
-2\pi i \d({\cal C})$, where $\d({\cal C})$ has been
introduced to denote the delta-two-form supported on the
critical circle ${\cal C}$, we obtain
\be
\varpi=12\pi \delta({\cal C}) d\phi.
\eel varpi

Let us conclude that the topological Wess-Zumino term may be
replaced by the sum of a local term and a
 topological term
supported on the set  ${\cal C}$ of antidiagonal matrices:
\begin{eqnarray} \label{Stop}
WZ(g)=\frac{ k }{4\pi} tr \int_{\Sigma}(d \gd \gdi\wedge d\gu \gui )+
S_\d(g),
\nonumber \\
S_\d(g)=\frac{ k }{12\pi}\int_{B}\varpi=  k \int_{B} \delta({\cal C}) d\phi.
\end{eqnarray}
The new topological term $S_\d(g)$ depends exclusively on
the positions of the points where $\Sigma$ intersects ${\cal C}$.
In particular, it vanishes if $\Sigma$ belongs to the part of
the group which admits the Gauss decomposition.

In Section 2 we showed that the local term of (\ref{Stop}) fits nicely
into the formalism of Poisson $\sigma$-models. Coupling of such
a model to the   topological term $S_\d$ is  subject of
Appendix C.

\section[]{Poisson $\sigma$-Model Coupled to a Topological Term
and \\  Quantum States for $SU(2)$-GWZW}
\setcounter{equation}{0}
Within this last Appendix we pursue the following three
goals: First we investigate
the change of a Poisson $\sigma$-model
\be
S_{{\cal P}}(X, A)=\int_{M}\left( A_{i \nu} \frac{\partial
X^i}{\partial x^\mu}  +
\frac{1}{2} {\cal P}^{ij} A_{i \mu} A_{j \nu}\right) dx^{\mu} \wedge
dx^{\nu} \,
\eel defSP
under  the addition of a topological term:
\be
S(X, A)=S_{{\cal P}}(X, A)+S_{top}(X) \, .
\eel S+S
Here $S_{top}(X)$ is supposed to be given by some closed three-form
$\omega_{top}$,
\be
S_{top}(X)=\int_B  \omega_{top} \; , \quad \6 B = \mbox{Image $M$} \; ,
\eel SX
of (generically) nontrivial cohomology on the target space $N$ of the model
(cf.\ also Appendix B). To not spoil the symmetries of  \ry defSP ,
we further require $\o_{top}$ to be invariant under any transformation
generated by vector fields of the form ${\cal P}^{ij}\6_j$.
We will focus especially on the change in the
Hamiltonian structure that is induced by \ry SX .

Next we will specify the considerations to the GWZW model. In the main
text and the previous Appendix we have shown that
the (Hamiltonian) GWZW action \rz fofac may be rewritten {\em identically}
in the form \rz S+S with $\omega_{top}=\varpi$. However, an additional
complication arises due to the fact that the matrix-valued one-form
\be  \b \equiv \b_i \, dX^i := \gui d\gu  -  \gdi d\gd \, ,   \eel alph
which we used in the identification \rz A   for $A$, becomes singular
at the part of $G$ where the  Gauss decomposition breaks down.
The singular behavior of $A$ has to be taken into account in the
variation for the field equations, if we want to
 describe the GWZW model by means of \rz S+S globally.
We will show that the bulk of the quantum states obtained in the
main text remains unchanged by these modifications.
The considerations change only for states that have support on
 loops lying on   exceptional conjugacy classes in $G$.

Finally we will make the considerations more explicit for $G=SU(2)$
and compare the  resulting picture to the literature.

In the classical Hamiltonian formulation the term \rz SX contributes
only into a change of the symplectic structure of the field theory. With
\be \o_{top}=\frac{1}{6} \, \o^{(top)}_{ijk} \, dX^i\wedge dX^j\wedge dX^k
\eel varcor
the symplectic structure  takes the form
\be
\Omega^{field}(X, A)= \oint_{S^1} dA_{i1}(x^1) \wedge dX^i(x^1)
dx^1 + \O^{field}_{top} \label{OmXA} \ee
with the extra piece
\be \O^{field}_{top}=
\frac{1}{2} \,  \oint_{S^1} \o^{(top)}_{ijk}(X(x^1)) \,
\partial_1 X^i (x^1) \, dX^j(x^1) \wedge
dX^k(x^1) \,  dx^1 . \label{Omtop}
\ee
Note that as (resp.\ if) $\o_{top}$ is non-trivial in cohomology on the
target space,
$\Omega^{field}$ becomes non-trivial as well, i.e.\ globally
there will not exist any symplectic potential $\Theta^{field}$ such that
$\Omega^{field} = d \Theta^{field}$.

In the case  $N=G$ and $\o_{top}:=\varpi$ the symplectic forms
\rz OmXA and \rz psss in the main text coincide. Actually  $A_{i1}(x^1)$
and $X^i(x^1)$ are Darboux coordinates of the symplectic form $\O^{field}$
of the GWZW model. As $\Omega^{field}$ has non-trivial  cohomology
such Darboux coordinates cannot exist globally. The situation may be compared
to the one of a sphere with standard symplectic form $\O=\sin \vartheta
d\vartheta  \wedge d\varphi$; trying to extend the local Darboux coordinates
$cos \vartheta$ and $\varphi$ as far as possible, one finds (again in a
distributional sense) $\O=
d\left(\cos \vartheta d\varphi \right) +
 2\pi\d^2(\mbox{'south pole'})- 2\pi\d^2(\mbox{'north pole'})$. Here we used
$d(d\varphi)=\sum_{poles}2\pi\d^2(pole)$, resulting from the breakdown of
$d\varphi$ as a coordinate differential at the poles while it still
remains a regular one-form in a distributional sense. By the way,
one may  infer eq.\ \rz psss  also from (\ref{OmXA},\ref{Omtop}):
Just replace the coordinate basis
$dX^i$ by the left-invariant basis $dg g^{-1}$ and note that
  $d(pdg g^{-1})=dpdg g^{-1} +
p(dg g^{-1})^2$ has to be substituted for  $d(A_{i1} dX^i) =dA_{i1} dX^i$.

The classical Gauss law \ry consXA ,
on the other hand, remains unaltered by the addition of a term \rz SX to the
action. Indeed the constraints  $G^i \approx 0$ emerge as the
coefficient of $A_{i0}$
within the action $S=S_{{\cal P}}+S_\d$ and  $S_\d$ does not depend on $A$.

Now let us turn to the quantum theory of the coupled model \ry S+S .
Again we go into an $X$-representation. In general 'wave functions'
may be regarded as section of a line bundle, the curvature of which
is the symplectic form \cite{Wood}. In the case that this line bundle
is trivial, i.e.\ when the symplectic form $\O^{field}$ allows for a
global symplectic potential,
one may choose a global non-vanishing section in the bundle.
The relative coefficient of any
other section with respect to the chosen one is then a
function, the wave function $\Psi[X]$. This
procedure is called trivialization of the line bundle.
In the case of prominent interest for us in which $\o_{top}$ and (thus)
$\O^{field}$ belong to some non-trivial cohomology class
 the quantum line bundle over the loop space will be non-trivial
\cite{GCarg}. Sections may be represented by
 functions $\Psi[X]$ then only within some local charts.

The $X^i$ may still be represented as multiplicative operators. However,
the change in the symplectic structure implies that one cannot
represent $A_{i1}$ as the derivative operators \rz A=dX any more.
Indeed the modification $\O^{field}_{top}$ preserves
commutativity of the $X^i$ as well as the commutation
relations between the $A_{i1}$ and  the $X^i$; however, the $A_{i1}$
do not commute among each other any longer.
The net result of the change in the symplectic
structure is that we have to add some $X$-dependent piece to
the operator representation of $A_{i1}$:
\be
A_{i1}=i \frac{\delta}{\delta X^i}+ \vartheta^{field}_i(X).
\eel Aopneu
The new quantity $\vartheta^{field}_i$ is a
symplectic potential to the non-trivial part  $\Omega^{field}_{top}$ of the
symplectic form, i.e.
\be
\vartheta^{field}_{top}=\oint \vartheta^{field}_i dX^i(x^1) dx^1
\eel F
is a  solution to the equation
\be
\Omega^{field}_{top}=d\vartheta^{field}_{top} \quad \mbox{(locally)} \, .
\eel OdF
$\vartheta^{field}_{top}$ is not unique and may be chosen in many ways.
If  $\o_{top}$ belongs to a  trivial cohomology class,
\rz OdF may be solved globally. Any choice for $\vartheta^{field}_i$ then
corresponds to the choice of a trivialization of this line bundle.
If, on the other hand,  $\o_{top}$ belongs to
some nontrivial cohomology class, we can speak about a solution to
\rz OdF only locally. Still any choice of a local potential
$\vartheta^{field}_{top}$  corresponds to a local
trivialization of the quantum line
bundle within some chart. Within the latter, quantum states may be
represented again as   ordinary functions $\Psi[X]$ on the loop space
and  \rz Aopneu  gives the corresponding operator representation of $A_{i1}$.

Let us finally write down the new quantum Gauss law. Within a
local chart on the loop space
it takes the form: \be i \left(\6 X^i +
{\cal P}^{ij}  \vartheta^{field}_j\right)\Psi =
{\cal P}^{ij} {\delta \0 \delta X^j} \Psi  \, . \eel conneu1
For  non-singular forms  $\vartheta^{field}_{top}$
these constraints yield a restriction
to functionals with  support on loops lying entirely within some
symplectic leaf again. (This holds true also for a singular
$\vartheta^{field}_j$,
as long as its contraction with the Poisson tensor ${\cal P}^{ij}$
vanishes).
To see this, just use the first $k$
coordinates $X^i$ to parameterize the symplectic leaves in any
considered region of $N$. Then \rz conneu1 yields
$\6 X^i \, \Psi =0$ for $i=1,...,k$. So, strictly speaking,
the physical wave functionals will
be  distributions that restrict the loops to lie entirely within
symplectic leaves. The remaining $n-k$ equations \rz conneu1 then determine
the form of $\Psi$ on each leaf.

Let us show this for trivial cohomology of the defining three-form
in \ry SX , i.e.\ for the special case that
\be \o_{top}=d\vartheta_{top} \eel triv
globally on $N$. Then $\vartheta^{field}_j =
\vartheta^{(top)}_{jk}\left(X(x^1)\right) \6_1 X^k(x^1)$ globally on the
phase space. To find the form of $\Psi$ on a given symplectic
leaf $\cal S$,
we multiply  \rz conneu1 for $i= k+1,\, ...\, , n$
by $\O_{li}$ from the left (cf.\ Eq.\ \ry sympl ).
The resulting equation can be integrated easily to yield:
\be \Psi = \Psi_0  \exp \left( i \int d^{-1} (\Omega + \vartheta_{top}) \right)
\eel loesung
where $\Psi_0$ is an integration constant, which, however, may depend
on the chosen symplectic leaf (and, if $\cal S$ is not simply connected,
also on the homotopy class of the argument loop of $\Psi$).
$\Psi_0$ may be regarded as the evaluation of $\Psi$ on
some reference loop on $\cal S$
and the phase is determined by the integration of
the two-form $\O+\vartheta_{top}$ over a two-surface that is enclosed between
the reference loop and the argument loop of $\Psi$. Independence of the choice
of the chosen two-surface  requires, e.g.\ for a simply connected $\cal S$:
\be \oint_\s \, \O+\vartheta_{top} = 2\pi n \;\; , \quad n \in Z \; ,\eel neu
for all two-cycles $\s \in \cal S$. This generalizes the
integrality condition \ry 63 , which corresponds to $\vartheta_{top} \equiv 0$.
\rz neu is a well-formulated condition, as
the invariance requirement for \rz SX  under the symmetries of \rz defSP
may be seen to imply that the {\em restriction} of $\vartheta_{top}$ onto
any symplectic leaf must be a closed two-form (while, certainly,
$\vartheta_{top}$
will not be closed in general on all of $N$).

For a truly {\em topological} term \rz SX equation \rz triv
holds only locally. Still \rz loesung provides
the local solution to the quantum constraints \rz conneu1 in the
space of loops on $\cal S$. However, as the form $\vartheta_{top}$ is not
defined globally on $\cal S$ in general, the global integrability
for \rz conneu1 does not have the simple form \ry neu . Instead the
use of various charts in the line bundle over the loop space will be
unavoidable to determine integrability of \rz conneu1 on a
leaf and thus the existence of a quantum state located on that leaf.
We will not study this problem in full generality here further.
Rather we will restrict our attention to the GWZW model in the following.

Everything that has been
written above applies to the GWZW model, too, except for one small
change: Actually, the correct Gauss
law for GWZW is not $G^i \approx 0$, but
\be \b_i G^i \approx 0 \, , \eel conneu2
 where the matrix-valued coefficients $\b_i$ have been defined in
\ry alph . To see this, we recall
that the constraints of the GWZW model, given first in
 Eq.\ \ry const , result from a variation for $\l \propto a_-$ within the
 action. According to
\rz A  $A_{i0}$ differs from (the components of) $\l$ by
$A_{i0}= tr \, \l \b_i$. So the correct GWZW Gauss law \rz const
may be rewritten as \ry conneu2 . For loops inside the
Gauss-decomposable region of $G$ this is equivalent to the old
form of the constraints $G^i =0$, since on that part of $G$ the
difference corresponds merely to a  change of
basis in $T^{\ast}G$.  However, as
$\b$ becomes singular at that lower dimensional part of $G$ where
the Gauss decomposition breaks down, the
constraints \rz conneu2 have  somewhat different implications than
$G^i =0$ in that region.

This consideration applies also to the quantum constraints;
we have to multiply \rz conneu1 by $\b_i$ from the left. The
 result is
\be  \left(\b_i \6 X^i + \b_i  {\cal P}^{ij}  \vartheta^{field}_j\right)\Psi +
i\b_i {\cal P}^{ij} {\delta \0 \delta X^j} \Psi =0 \, , \eel conneu3
or, equivalently,
\be
\left(\gui \6_1\gu - \gdi\6_1\gd + \b_i  {\cal P}^{ij}  \vartheta^{field}_j +
\frac{2\pi i}{ k } \frac{\delta}{\delta
h} h\right) \Psi[g_0, h] =0 \, .
\eel hgl2
The part $\b_i  {\cal P}^{ij}  \vartheta^{field}_j$, which may be rewritten
also as the insertion of the vector field $(2\pi/k) (\d / \d h)h$ into
the one-form $\vartheta^{field}_{top}=\vartheta^{field}_\d$,
is the new contribution from $S_\d$ that has been dropped in
the derivation of \ry hgl .

It is not difficult to see that for loops that lie at least partially
outside exceptional conjugacy classes ('critical region') in $G$ one
may solve \rz PsiX instead of
\rz conneu3 or \ry hgl2 . Indeed close to any part of the loop outside
the critical region we may use \rz conneu1 as the
quantum constraint, because \rz alph is invertible in that part of
$G$. But as argued above this restricts the loop to lie {\em entirely}
within a  symplectic leaf outside the critical region in $G$.
For {\em such} loops now  we may always choose
\be \vartheta^{field}_i \equiv 0   \, , \eel thetanull
as $\O^{field}_\d$ vanishes on that part of the phase space.
This justifies that in the main text we dropped the contributions from
$S_\d$ (as well as the multiplicative factor $\b_i$) and restricted
our attention to the solution of  \ry PsiX .
Also we had not to think of a non-trivial
quantum line bundle in this way. The main part of the states could be
obtained within one local trivialization of the line bundle, given by
\ry thetanull .

What has to be considered separately  only
are possible states that have support on loops
lying {\em entirely} within the critical region of $G$.
In this case the full quantum constraints \rz conneu3
have to be taken into account. It is also in this region of $G$, furthermore,
where the notion of symplectic leaves and conjugacy classes do not
coincide. From \rz hgl2 we learn that it is precisely the
modifications of \rz PsiX that restore the adjoint transformations as
symmetries on the quantum level. $\b$ diverges precisely where $\cal P$
vanishes so as to give rise to the finite contribution
$(\d/\d h) \, h$ in \ry hgl2 . As a result there will correspond at
most one quantum state to an exceptional conjugacy class, even if the
respective orbit splits into several (possibly in part integrable)
symplectic leaves.

Let us now specify our considerations to $G=SU(2)$. In particular we
want to determine all quantum states within our approach. For this
purpose let us first consider the splitting of $SU(2)\sim S^3$ into conjugacy
classes.  Parameterizing conjugacy classes by (cf.\ also  \ry parm )
\be \2 tr\, g = \mbox{Re}(z) = : \cos \theta = const \: , \quad
\theta  \in [0,\pi]\ ,
\eel pz
we find that, topologically speaking, these orbits
are two-spheres for $\theta \in \; (0,\pi)$ and
points for $\theta = 0, \pi \lra z = \pm 1$. Only one of the conjugacy
classes is 'exceptional'; it corresponds to $\theta = \pi/2 \lra tr g = 0$.
Parameterizing this critical $S^2$ by
polar coordinates $\phi$ and $\vartheta := \arccos \mbox{Im}(z)$, the
part $\cal C$
of $SU(2)$ on which the Gauss decomposition is not applicable is
identified with the equator $\vartheta = \pi/2$ of this two-sphere.

So the picture we obtain is that $N = S^3$ is foliated into
two-spheres except for its 'poles' $z=\pm 1$.
The 'equator' of the three-sphere, itself
an $S^2$,  is what we called an exceptional conjugacy class.
The  equator ${\cal C} \sim S^1$ of this $S^2$ is precisely
the subset of $N=G$ where the Gauss decomposition breaks down and,
correspondingly, where  the support of $\o_{top}=\varpi$ lies. The
exceptional conjugacy class
splits into several symplectic leaves: the Northern part of the
$S^2$, its Southern part, and the points of the equator
${\cal C}$, where ${\cal P}$ vanishes. According to our general
considerations above, this splitting is, however, irrelevant; there
will correspond at most  one quantum state to the exceptional
conjugacy class.

On the other hand there  corresponds precisely one quantum state to
any integral (non-exceptional) conjugacy class, as all of these orbits are
simply connected. So let us  evaluate the integrality condition \rz 63 for
the non-exceptional conjugacy classes in $SU(2)$. From \rz omega
we find that in the coordinates \rz parm
\be
   \Omega = {i k \over 2\pi}{dz\over z} \wedge d\phi \, .
\eel omonsu2
In the parameterization \rz pz for  the adjoint orbits this yields for
the integral of $\Omega$ over the respective two-spheres
\be
  \int_{S^2}\Omega = \left\{ \begin{array}{r@{\quad , \quad}l}
2 k \theta & \theta \in \, [0,\pi/2) \\ 2 k (\pi -\theta) &  \theta \in
\;\; (\pi/2,\pi] \end{array} \right. \eel volume
Here we have taken into account that the imaginary part Im$(z)$
of $z$ runs
only from $-\sin \theta$ to $+\sin \theta$ since $|z| \le 1$.

For the critical orbit at $\theta = \pi/2$ the symplectic volume
\rz volume becomes ill-defined. This comes as no surprise. Here obviously
the choice \rz thetanull does not apply for all loops on the
critical conjugacy class. Still the correct integrability
condition may be guessed  from a simple limiting procedure: From
\rz volume we obtain
\be
\lim_{\theta \to \pi/2}\; \, \int_{S^2}\Omega =  k\pi .
\eel aa
It is
plausible to assume that the critical orbit will carry a quantum state,
iff again \rz aa is an integer multiple of $2\pi$ (cf.\ Eq.\ \ry 63 ).

In fact, one can prove that this is indeed correct.
To do so one might use two charts in the quantum line bundle.
First \ry thetanull , which works for all loops that do  not intersect
the equator  ${\cal C}$ of the critical conjugacy class. And second,
\be \vartheta_\d := {k \0 2 \pi} \left({1 \0 z} +i\right) dz \wedge
d\phi \quad \longrightarrow \quad \vartheta^{field}_i =
{k \0 2 \pi} \left({1 \0 z} +i\right) (dz \, \6_1 \phi - \6_1 z \,
d\phi)  \, .
\eel chart2
This second chart is applicable to all loops on the critical
conjugacy class that do not touch its 'pole'   $z=-i$. The solution to
the full quantum constraints \rz conneu3 has again the form \rz loesung
within the respective domain of definition of the two charts. Now one
might regard the value of the wave functional in {\em both} charts
for two small loops close to the pole $z=-i$, one of which
with winding number one around this pole, the other one with
winding number zero. In the first chart
continuity of the wave function implies that the wave functional will have
basically the same value for both loops. In the second chart the two loops
are separated from each other by a two-surface that encloses basically
all of the critical $S^2$ (since in this chart the first loop may not be
transformed into the second one through the pole $z=-i$, but instead one
has to move through the other pole $z=i$); this gives a relative phase factor
of the wave functions in this chart that may be determined by means of
\ry loesung . The corresponding phase need, however, {\em not}
be a multiple of $2\pi$. Instead, the result of chart two
has to coincide with the result of chart one only {\em after} taking into
account the transition functions between the two charts. (Note that
both loops lie in both charts). In fact, for the first loop one
picks up a nontrivial contribution to the integrality  condition
from there. Further details shall be left to the reader.
In any case the result coincides with the one obtained from the limit above.
So one finds that there exists
a quantum state with support on the critical orbit $\theta = \pi/2$
for even values of $k$ and no such a state for odd values of $k$.

Let us remark here that in the latter case {\em all} $\,$
'physical' quantum states, i.e.\ all
states in the kernel of the quantum constraints, may be
described within just one chart of the quantum line bundle (as e.g.\ by
\ry thetanull ). So, the restriction to physical states
may yield the originally
non-trivial quantum line bundle of a coupled model
\rz S+S to become  effectively trivial.

Summing up the results for $G=SU(2)$, we conclude that
the integral orbits (i.e.\ the
orbits allowing for  nontrivial quantum states of the $SU(2)$-GWZW model)
are given by $\theta = n\pi/ k $, $n=0,1,... k$.

Now we want to compare this result with the current literature.
According to \cite{ADW}, there are two different pictures for
the space of states of the GWZW model. (In \cite{ADW} they
consider partition functions of the WZW model. However, these
two issues may be  related using results of \cite{GK}.) The first
picture eventually coincides with our answer. The second one
suggests the finite renormalization $k\rightarrow k+2$. In this case
the integral orbits are characterized by $\theta= n\pi/(k+2), \
n=0,\dots , k+2$. However, in {\em this}
picture the singular orbits with $n=0$ and
$n=k+2$, corresponding to the central elements $\pm I\in SU(2)$,
should be excluded. In \cite{ADW} it is proved that the two pictures
are equivalent. However, it would be interesting to establish this
equivalence in the language of Poisson $\sigma$-models. One motivation
is to compare the results with the similar formalism \cite{Blau}.
Also, it seems to be easier to handle the spectrum of the model
in the second picture.

\end{appendix}

\section*{Acknowledgements}

We are grateful to M.Blau, K.Gawedzki and R.Jackiw
for the interest in this work and useful comments. A.A. thanks the
Erwin Schr\"{o}dinger Institute (Vienna, Austria) for hospitality
during the period when this work was initiated.


\begin{thebibliography}{99}

\bibitem{AER} D.Amati, S.Elitzur and E.Rabinovici, Nucl.\ Phys.\ {\bf B
    418} (1994) 45.
\\
D.Cangemi, R.Jackiw, Phys.\ Rev.\ {\bf D 50}/6 (1994) 3913.
\\
S.Shabanov, P.Schaller and T.Strobl, unpubl.\ collaboration, Nov.\ 1993.
\\ Cf.\ also reference \cite{Th} below.

\bibitem{JG} J.Goldstone and R.Jackiw, Phys.\ Lett.\ {\bf B 74} (1978) 81.

\bibitem{FRS} A.G.Izergin, V.E.Korepin, M.A.Semenov-Tyan-Shanskii,
L.D.Faddeev,
Teor.\ Mat.\ Phys.\ {\bf 38}/3 (1979) 1.

\bibitem{Th} P.Schaller and T.Strobl, \\
{\em Quantization of Field Theories Generalizing Gravity-Yang-Mills
      Systems on the Cylinder}, In:  {\bf LNP}  {\bf 436},
      p.\ 98, 'Integrable Models and Strings', eds.\ A.Alekseev et al,
      Springer 1994 or gr-qc/9406027; \\
   {\em Mod.\ Phys.\ Letts.} A9 (1994) 3129; \\
      cf.\ also {\em Poisson $\sigma$-models: A generalization of
        ...}, to appear in Proc.\ of  'Conference on Integrable
     Systems', Dubna 1994  or hep-th/9411163.
\bibitem{Wess} J.Wess and B.Zumino, Phys.\ Lett.\ {\bf B 37}/1 (1971) 95.

\bibitem{FSh} L.D.Faddeev and S.L.Shatashvili, Teor.\ Math.\ Phys.\ {\bf 60}
(1985) 770.
\\
L.Faddeev, Phys. \ Lett.\ {\bf B 145} (1984) 81.

\bibitem{Sp} M.Spiegelglas, Phys.\ Lett.\ {\bf B 247}/1 (1990) 36.

\bibitem{Blau} M.Blau and G.Thompson, Nucl.\ Phys.\ {\bf B 408}
(1993) 345.
\\
A.Gerasimov, {\em Localization in GWZW and Verlinde formula},
Uppsala preprint and hep-th/9305090.

\bibitem{GK} K.Gawedzki and  A.Kupiainen, Nucl.\ Phys.\ {\bf B 320} (1989)
625.

\bibitem{AS} A.Alekseev and S.Shatashvili, Nucl.\ Phys.\ {\bf B 323}
(1989) 719.
\\
M.Bershadskii and  H.Ooguri, Commun.\ Math.\ Phys.\ {\bf 126}
(1989) 49.
\\
A.Gerasimov, A.Marshakov, A.Morozov, M.Olshanetsky,
S.Shatashvili, Int.\ J.\ Mod.\ Phys.\ {\bf A 5}/13 (1990) 2495.

\bibitem{Th2} T.Strobl, {\em Phys.\ Rev.} {\bf D 50}/12 (1994)  7356.
\\
T.Kloesch and T.Strobl, {\em All Global Solutions of 1+1 Dimensional
Gravity}, preprint.


\bibitem{Sem} M.A.Semenov-Tian-Shansky,
{\em Dressing transformations and Poisson-Lie group
actions,} in:
Publ.\ RIMS, Kyoto University {\bf 21}/6 (1985) 1237.

\bibitem{Nov} S.P.Novikov, Dokl.\  Acad.\  Nauk\ SSSR {\bf 260}
(1981) 31.
\\
S.P.Novikov, Usp.\  Mat.\ Nauk\ {\bf 37} (1982) 3.

\bibitem{JW} S.Deser, R.Jackiw, S.Templeton, Ann.\ Phys.\ {\bf 140}
(1982) 372.
\\
E.Witten, Nucl.\ Phys.\ {\bf B 223} (1983) 422.

\bibitem{Wood}  N.M.J.Woodhouse,  {\em Geometric
Quantization}, second Edition 1992, Clarendon Press, Oxford.

\bibitem{NR} H.B.Nielsen and  D.Rohrlich, Nucl.\ Phys.\ {\bf B 299} (1988)
471.
\\
 A.Alekseev, L.Faddeev and  S.Shatashvili,
 J.\ Geom.\ Phys.\ {\bf 5}/3 (1989) 391.


\bibitem{AM}  A.Yu.Alekseev and  A.Z.Malkin, Commun.\ Math.\ Phys.\ {\bf
162} (1994) 147.

\bibitem{AT} A.Alekseev and  I.Todorov, Nucl.\ Phys.\ {\bf B 421}
(1994) 413.


\bibitem{WLu} J.H.Lu and  A.Weinstein,
J.\ Diff.\ Geom.\ {\bf 31}  (1990) 501.

\bibitem{FRT} L.D.Faddeev, N.Yu.Reshetikhin and L.A.Takhtadjan,
Leningrad Math.\ J.\ {\bf 1} (1989) 178.

\bibitem{BB} B.Block, Phys.\ Lett.\ {\bf B 233} (1989) 359.
\\
L.Faddeev, Commun.\ Math.\ Phys. \ {\bf 132} (1990) 131.
\\
A.Alekseev and S.Shatashvili, Commun.\ Math.\ Phys.\ {\bf 133} (1990) 353.
\\
F.Falceto and K.Gawedzki,
{\em On quantum group symmetries of conformal field theories,}
in: Proc.\ XXth Int.\ Conf.\ on Differential geometric methods in
theoretical physics, New York, 1991;
\\
Bur-sur-Yvette preprint IHES/P/91/59.
\\
J.Balog, L.Dabrowski, L.Feh\'{e}r, Phys.\ Lett.\ {\bf B 244}
(1990) 227.
\\
M.Chu, P.Goddard, I.Halliday, D.Olive, A.Schwimmer,
Phys.\ Lett.\ {\bf B 266} (1991) 71.

\bibitem{GCarg} K.Gawedzki,
{\em Topological actions in two-dimensional quantum field theory,} in:
Non-perturbative quantum field theory, pp.101-142, eds.\ G.t'Hooft et al,
Plenum Press, New York, London 1988.

\bibitem{ADW} S.Axelrod, S.Della Pietra, E.Witten,
J.\ Diff.\ Geom.\ {\bf 33} (1991) 787.

\end{thebibliography}
\end{document}